\documentclass{article}

\usepackage{cite}
\usepackage{graphicx}
\usepackage{amsmath}

\def\scaleval{0.33}
\newcommand{\odiffx}[1]{\frac{d #1}{d x}}
\newcommand{\pdifftz}[1]{\frac{\partial #1 \hfill}{\partial t_0}}
\def\ulin{u_{\rm lin}}
\def\rlin{\rho_{\rm lin}}
\def\uhm{{\hat{u}_m}}
\def\uhn{{\hat{u}_n}}
\def\rhl{{\hat{\rho}_\ell}}

\def\rhn{{\hat{\rho}_n}}

\begin{document}

\title{Phase distortion mechanisms in\\linear beam vacuum devices}
\author{John G.~W\"ohlbier}
\maketitle

\begin{abstract}
  The mechanism for phase distortion in linear beam vacuum devices is
  identified in the simplest system that pertains to such
  devices, the force-free drifting electron beam. We show that
  the dominant cause of phase distortion in a force-free drifting beam
  is the inverse dependence of arrival time on velocity
  for an electron, i.e., the ``$1/u$ nonlinearity,'' and that a
  secondary
  cause of phase distortion is the nonlinearity of the velocity
  modulation. We claim that this is the mechanism for phase distortion
  in all linear beam vacuum devices, and provide evidence from a
  traveling wave tube calculation.
  Finally, we discuss the force-free
  drifting beam example in the context of the ``self-intermodulation''
  description of phase distortion recently
  described in Refs.~[J.~W\"ohlbier and J.~Booske, Phys.~Rev.~E,
  vol.~69, 2004, 066502], [J.~W\"ohlbier and J.~Booske, IEEE
  Trans.~Elec.~Devices, To appear.]
\end{abstract}

We recently reported on mechanisms for phase distortion in linear beam
vacuum devices~\cite{wohlbier:ampm04, wohlbier:nscw05}. The mechanism
for phase distortion was reported there as a
``self-intermodulation'' process, where harmonic beam distortions
interacted with the fundamental beam modulation to produce a phase
shift at the fundamental.
While this frequency domain view of phase distortion is a
useful one, we felt that a corresponding time domain view of phase
distortion would be a useful contribution to the overall
understanding of phase distortion.
In this note we consider
the simplest possible system that pertains to linear beam electron
devices, a space charge free (``force-free'') drifting electron beam.
Indeed the physical mechanism for phase distortion exists in the 
force-free drifting beam, and hence the system provides the most lucid
example in which to study phase distortion.

A 1-d force-free drifting beam is described by Burger's equation
for the beam velocity $u$,
\begin{eqnarray}
  \label{eq:burgers}
  u_t + u u_x & = & 0,
\end{eqnarray}
where $x$ is space, $t$ is time, and the subscripts indicate partial
derivatives.
To apply Eq.~(\ref{eq:burgers}) to a
klystron where the space charge force is negligible, for example, one
sets the boundary value of $u$ as the sum
of a dc beam
velocity $u_0$ and a sinusoidal perturbation
due to the cavity modulation. Assuming that the dc velocity is
normalized to~$1$, this is written
\begin{eqnarray}
  \label{eq:velocity_mod}
  u(0,t) & = & 1 + \epsilon \sin{\omega t}.
\end{eqnarray}
Given the solution to Eq.~(\ref{eq:burgers}) with boundary
condition~(\ref{eq:velocity_mod}), the beam
density evolution is obtained by solving the continuity equation
\begin{eqnarray}
  \label{eq:continuity}
  \rho_t + (\rho u)_x & = & 0
\end{eqnarray}
for an appropriate density boundary condition.

The force-free Burger's equation (\ref{eq:burgers}) implies that the
velocity of an electron (fluid element) does not change, i.e., the
$(x,t)$ trajectories, or
characteristics, of Eq.~(\ref{eq:burgers}) are straight lines with
slopes
determined by the boundary data. In vacuum electronics
the density bunching that results from the sinusoidal velocity
modulation is called ``ballistic bunching.''
Initially
we will consider inputs for which there is no electron
overtaking, and inputs such that electron overtaking occurs will be
considered later.

In principle the force-free drifting electron beam can be
solved exactly prior to electron overtaking, although closed form
analytic solutions need to be written in terms of infinite
series~\cite{lau:tik00}.
Equation (\ref{eq:burgers}) is solved using
the method of characteristics. The method of characteristics involves
changing the equations to Lagrangian independent coordinates (material
coordinates), solving the equations
in Lagrangian coordinates, and changing the solution back to Eulerian
coordinates.
For a fluid element that crosses $x=0$ at time $t_0$ we write the
transformation from Lagrangian to Eulerian coordinates as the time
function $t(x,t_0)$, i.e., the time fluid element $t_0$ arrives at
$x$, with $t(0,t_0) = t_0$. Since the velocity is a
constant for each fluid element, the velocity solution in
Lagrangian coordinates is
\begin{eqnarray}
  \label{eq:u_lc}
  u(x,t_0) & = & 1 + \epsilon \sin{\omega t_0}.
\end{eqnarray}
That is, independent of location $x$, the velocity of an electron is
set by the time $t_0$ at which it crosses $x=0$.
Since the electron orbits are straight line trajectories
in $(x,t)$ space, we can infer the solution for the function
$t(x,t_0)$
\begin{eqnarray}
  \label{eq:tfn_1}
  t(x,t_0) & = & t_0 + \frac{x}{u(x,t_0)}\\
  \label{eq:tfn_2}
  & = & t_0 + \frac{x}{1 + \epsilon \sin{\omega t_0}}.
\end{eqnarray}
To solve for $u(x,t)$ one needs to invert the function $t(x,t_0)$,
i.e., calculate $t_0(x,t)$, and substitute it into
Eq.~(\ref{eq:u_lc}). Since we restrict, for now, our attention to input
levels such
that no electron overtaking occurs, i.e., the characteristics do not
cross, Eq.~(\ref{eq:tfn_2}) can be inverted. Unfortunately
$t(x,t_0)$ in Eq.~(\ref{eq:tfn_2}) is a transcendental function of
$t_0$, so an analytic inverse
needs to be expressed in terms of an infinite series.

For phase distortion we are ultimately interested in the beam current
and beam density, since they linearly drive circuit or cavity
fields.
The continuity equation in Lagrangian coordinates
is given by~\cite{wohlbier:muse02,lau:tik00}
\begin{eqnarray}
  \rho(x,t_0) & = & \frac{\rho(0,t_0) u(0,t_0)}{\left| \frac{\partial
        t \hfill}{\partial t_0} \right| u(x,t_0) }\\
  \label{eq:lagrangian_continuity}
  & = &  \frac{\rho(0,t_0)}{\left| \frac{\partial
        t \hfill}{\partial t_0} \right|}
\end{eqnarray}
where the last equality comes from the fact that
$u(0,t_0) = u(x,t_0)$ since we are
considering a force-free drifting beam. For the density solution in
Eulerian coordinates one composes the density in Lagrangian
coordinates (\ref{eq:lagrangian_continuity}) with the mapping from
Eulerian to Lagrangian coordinates $t_0(x,t)$.

The factor $\partial t / \partial t_0$ is the Jacobian of the
transformation from Lagrangian to Eulerian coordinates, and it
quantifies the amount of stretching or compressing in time the
electron beam
undergoes. That is, for a fluid element entering the system of time
length $\delta t_0$, it will occupy a time length of $\delta t =
(\partial t(x,t_0)/\partial t_0) \delta t_0$ downstream at position
$x$. For purposes of explaining phase distortion it will be convenient
to write the Jacobian in terms of derivatives with respect to both
$t_0$ and $u$, i.e.,\footnote{We use the function $\hat t(x,t_0,u)$,
  cf.~Eq.~(\ref{eq:tfn_1}),
  to formally differentiate from
  the function $t(x,t_0)$, cf.~Eq.~(\ref{eq:tfn_2}),
  so that the left hand side
  of Eq.~(\ref{eq:jacobian})
  and the first
  term on the right hand side of Eq.~(\ref{eq:jacobian}) are not considered
  equal. We did not use the
  hat ($\hat{\rule{2mm}{0mm}}$)
  notation in Eqs.~(\ref{eq:tfn_1}) and (\ref{eq:tfn_2}) where there was no
  chance of confusion, and we will drop the
  notation for the remainder of the paper.}
\begin{eqnarray}
  \label{eq:jacobian}
  \pdifftz{t} & = & \pdifftz{\hat t} + \frac{\partial
    \hat t}{\partial u \hfill} \pdifftz{u}.
\end{eqnarray}

\begin{figure}[htbp]
  \centering
  \includegraphics*[scale=\scaleval]{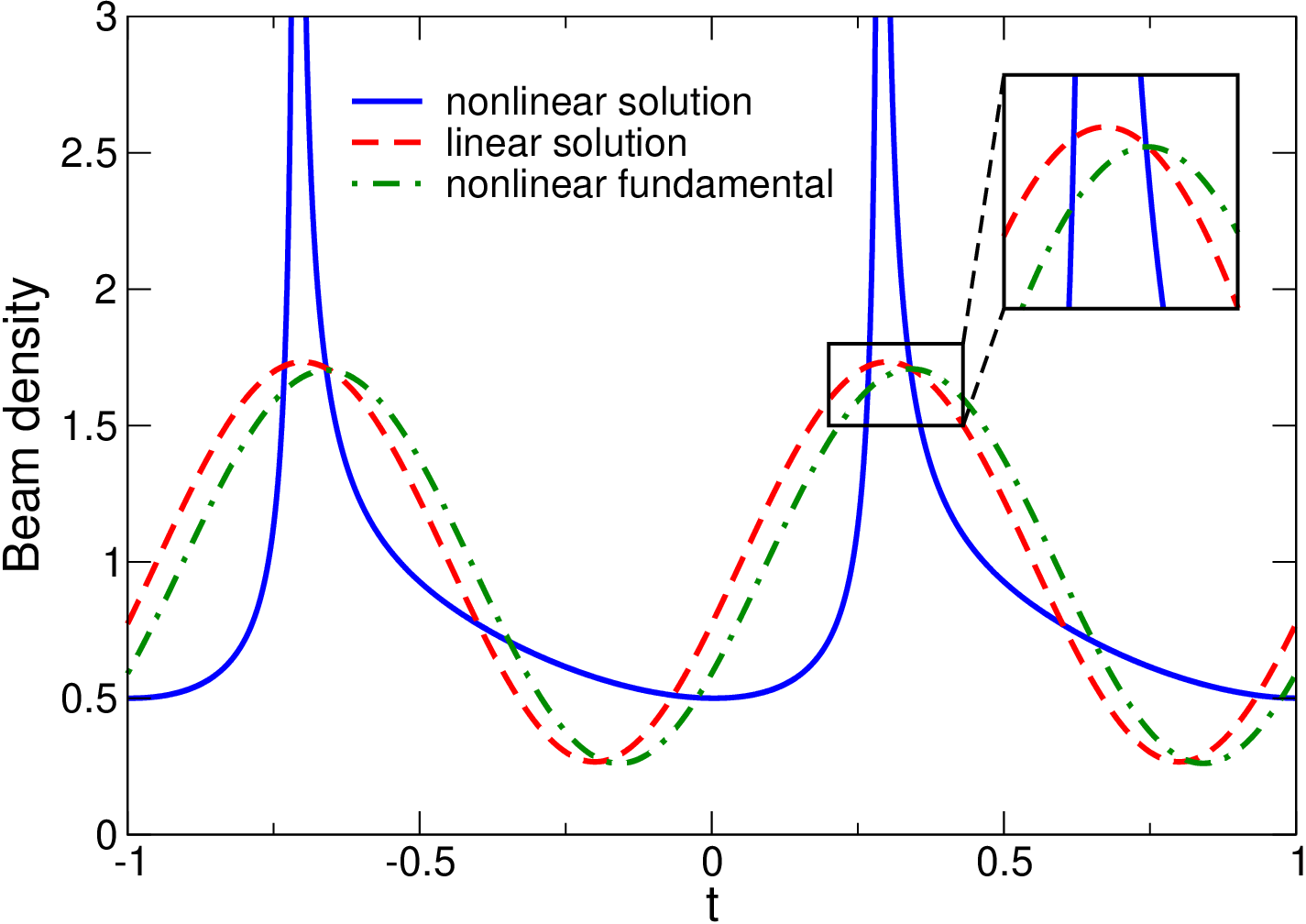}
  \caption{Nonlinear, linear, and fundamental
    component of nonlinear density solution versus $t$ at $x=0.3$ for
    two periods.}
  \label{fig:nl_l_fc}
\end{figure}

In Fig.~\ref{fig:nl_l_fc} we show density
solutions versus time, where
a root finder was used to invert Eq.~(\ref{eq:tfn_2}).
The results are for the velocity modulation (\ref{eq:velocity_mod})
with $\omega = 2\pi$ and a uniform density boundary condition
$\rho(0,t_0) = 1$. To
accentuate the effect of phase distortion we use a strong velocity
modulation, $\epsilon = 0.389$, and observe the output at $x =
0.3$. There is no electron overtaking at $x$ if the Jacobian stays
greater than zero, i.e., when
\begin{eqnarray}
  \label{eq:condition}
  x \frac{\epsilon \omega \cos \omega t_0}{\left(1 + \epsilon \sin
      \omega t_0 \right)^2} & \le & 1
\end{eqnarray}
for all $t_0$.
In the present case the maximum value of the left hand side of
Eq.~(\ref{eq:condition}) is equal to $0.9992$ with $x = 0.3$. The linear
velocity and density solutions are solutions to the (normalized)
linear equations
\begin{eqnarray}
  u_t + u _x & = & 0\\
  \rho_t + \rho_x + u_x & = & 0
\end{eqnarray}
and, for the same boundary data as used for the nonlinear problem, are
given by
\begin{eqnarray}
  \label{eq:u_linear}
  \ulin(t,x) & = & 1 + \epsilon \sin \omega (t-x)\\
  \label{eq:rho_linear}
  \rlin(t,x) & = & 1 + x \epsilon \omega \cos \omega (t-x).
\end{eqnarray}

In Fig.~\ref{fig:nl_l_fc}
the nonlinear density solution, the linear
density solution,
and the fundamental component of the nonlinear density solution
obtained by using a fast Fourier transform are shown.
The phase shift between the linear density
solution and the fundamental component of the nonlinear density
solution is the ``phase distortion'' of the nonlinear solution.
It is clear from the figure that the reason
for the phase shift of the fundamental component of the nonlinear
density solution from the linear solution is because the density is
larger ``on average'' behind (in time) the density peak (early times
are to the left). Therefore, to explain phase distortion one must
explain the reason for the asymmetry in the density about the density
peak. (Also note that for this value of $\epsilon$ the nonlinear
solution shown in Fig.~\ref{fig:nl_l_fc} shows ``amplitude
distortion'' in that the fundamental component of the density is
smaller than the predicted linear density amplitude.)

The reason for the the higher density behind the peak comes
predominantly from
the ``$1/u$ nonlinearity'' in the arrival time function
(\ref{eq:tfn_1}), and secondarily from the nonlinearity of the
$\epsilon\sin\omega t_0$
velocity modulation. It is intuitive that about a larger initial
velocity $u$, a given 
deviation $\pm\delta u$ will result in a smaller $\mp\delta t$ at $x$
than the same $\pm\delta u$ would have on $\mp\delta t$ for a
smaller initial velocity $u$
[a fluid
element with velocity $u \pm \delta u$ arrives earlier (later), at time
$t \mp \delta t$, than a fluid element with velocity $u$ which arrives
at $t$].
This intuition is of course borne out by
the derivative
\begin{eqnarray}
  \frac{\partial t}{\partial u} & = & - \frac{x}{u^2}
\end{eqnarray}
evaluated at large and small values of velocity $u$. This dependence
causes phase distortion in that relatively slower fluid elements
stretch and compress differently than the relatively faster fluid
elements. That is, the size of density fluctuations at a fluid element
depends on its initial velocity, with smaller initial velocities
having larger fluctuations.
This effect is best demonstrated with an example where more specific
points regarding the interplay of the $1/u$ nonlinearity and the
nonlinearity of the velocity modulation may be emphasized.

\begin{figure}[htbp]
  \centering
  \includegraphics*[scale=\scaleval]{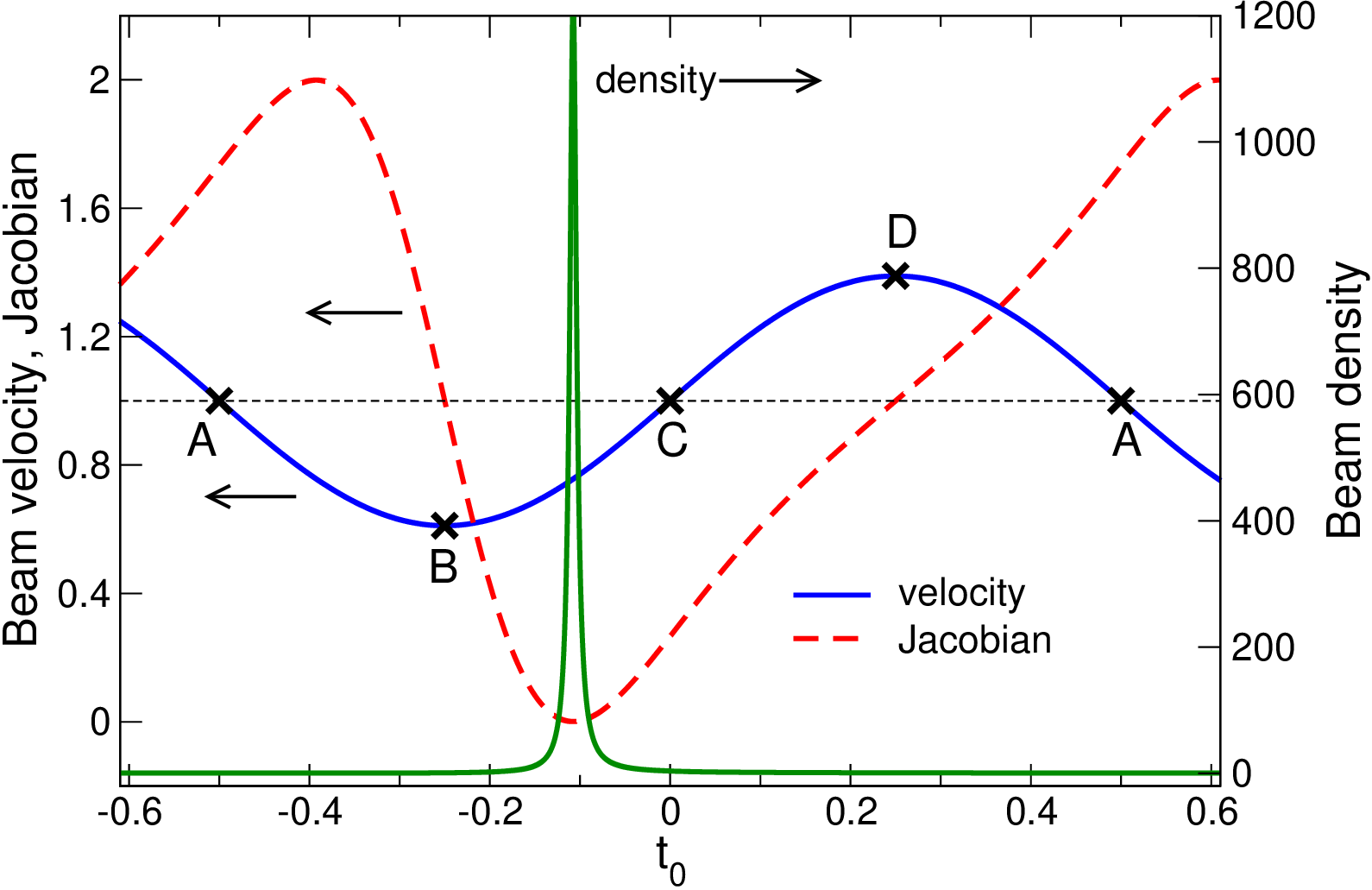}
  \caption{Beam velocity, beam density, and Jacobian
    versus fluid element label~$t_0$ at $x=0.3$. That is, the value of
    velocity, density and Jacobian that the fluid element that crossed
    $x=0$ at time $t_0$ will have when it arrives at $x=0.3$. Results
    show slightly more than one period in $t_0$.}
  \label{fig:v_j_d_vt0}
\end{figure}

In Fig.~\ref{fig:v_j_d_vt0} we show the velocity modulation,
the Jacobian, and the density versus fluid element $t_0$ at $x=0.3$.
Note that since $u(x,t_0)=u(0,t_0)$ the velocity versus fluid element
label $t_0$ is also the initial velocity modulation at $x=0$ versus
fluid element label $t_0$.
In the region B--D the beam is bunched because fluid elements entering
at any time are given a faster initial velocity than the fluid
elements entering just prior to them, where again we assume for the
time being that modulations are not strong enough to cause electron
overtaking. It is instructive to look at the expression for the Jacobian,
\begin{eqnarray}
  \label{eq:jacobian2}
  \pdifftz{t} & = & 1 + \frac{\partial t}{\partial u}
  \frac{\partial u\hfill}{\partial t_0},
\end{eqnarray}
and determine, for example, the fluid element at which the density
will be maximum.
The Jacobian is inversely proportional to density, and hence the
maximum density occurs the minimum Jacobian; for a given $\delta t_0$
the minimum $\delta t$ corresponds to the maximum fluid compression.
Consider the two derivatives on the right hand side of
Eq.~(\ref{eq:jacobian2}).
As described above, due to the inverse dependence of $t$ on $u$,
for a given $\delta u$ fluid elements with relatively larger initial
velocities in B--D
will be compressed less than fluid elements with relatively smaller
initial velocities.
That is, $\partial t/\partial u$ is largest negative at B and
smallest negative at D.
Furthermore, for the sinusoidal velocity modulation the
change in velocity for a given $\delta t_0$, $\partial u/\partial
t_0$, is zero at point B and increases to a maximum at point C. The
minimum Jacobian will occur at a fluid element between points B and C
such that $(\partial t/\partial u) (\partial u/\partial t_0)$ is the
largest negative.
The value of $t_0$ for which this happens can be computed by
setting the derivative of the Jacobian with respect to $t_0$ to zero,
and solving for $t_0$. For the linear solution point C is the fluid
element around which the density is maximum.

\begin{figure}[htbp]
  \centering
  \includegraphics*[scale=\scaleval]{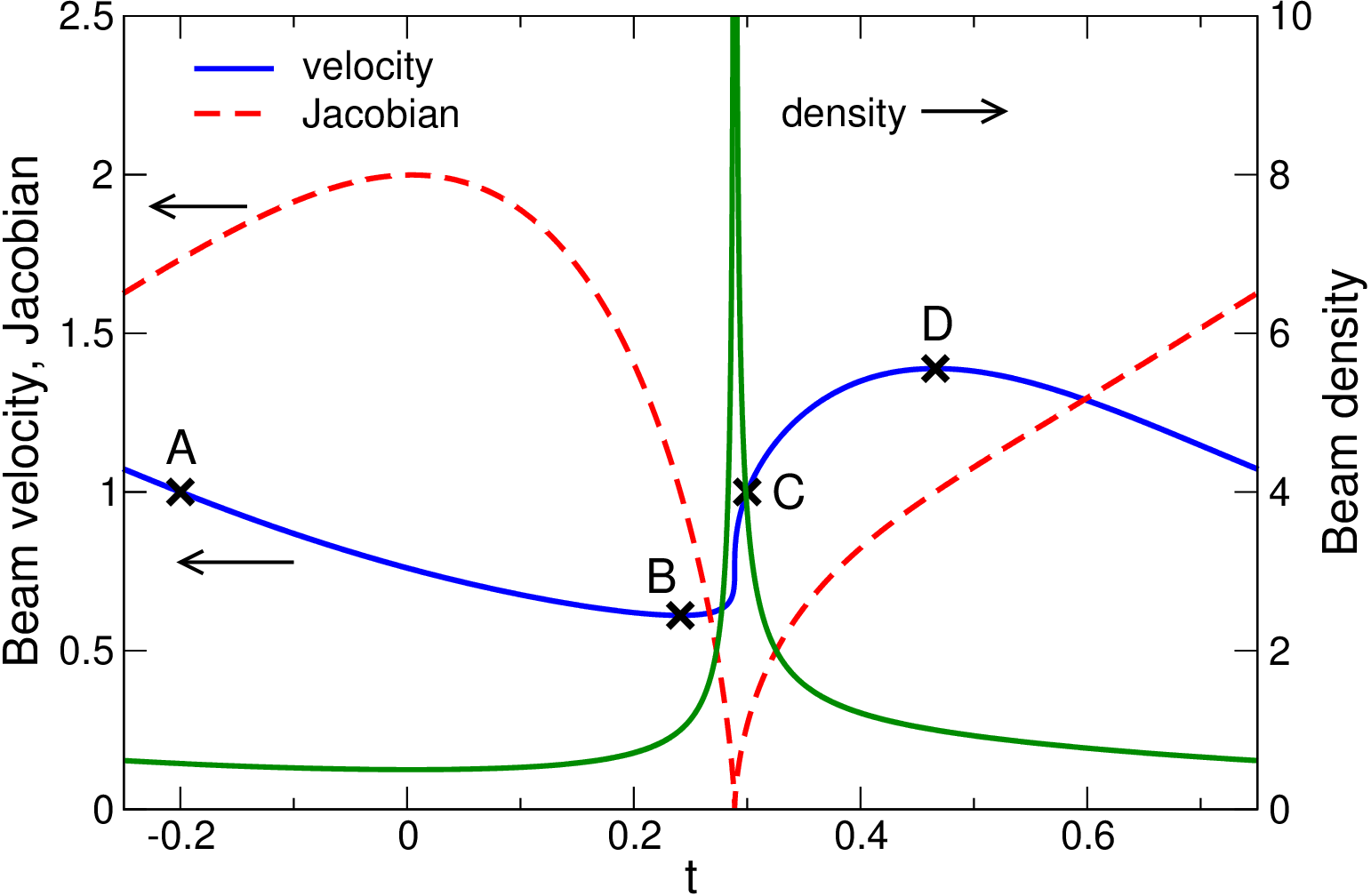}
  \caption{Beam velocity, density, and Jacobian versus
    $t$ at $x=0.3$
    for one period of $t$. Results
    are those of Fig.~\ref{fig:v_j_d_vt0} composed with $t_0(x,t)$.
    shown in Fig.~\ref{fig:t0_map}.}
  \label{fig:v_j_d_vt}
\end{figure}

\begin{figure}[htbp]
  \centering
  \includegraphics*[scale=\scaleval]{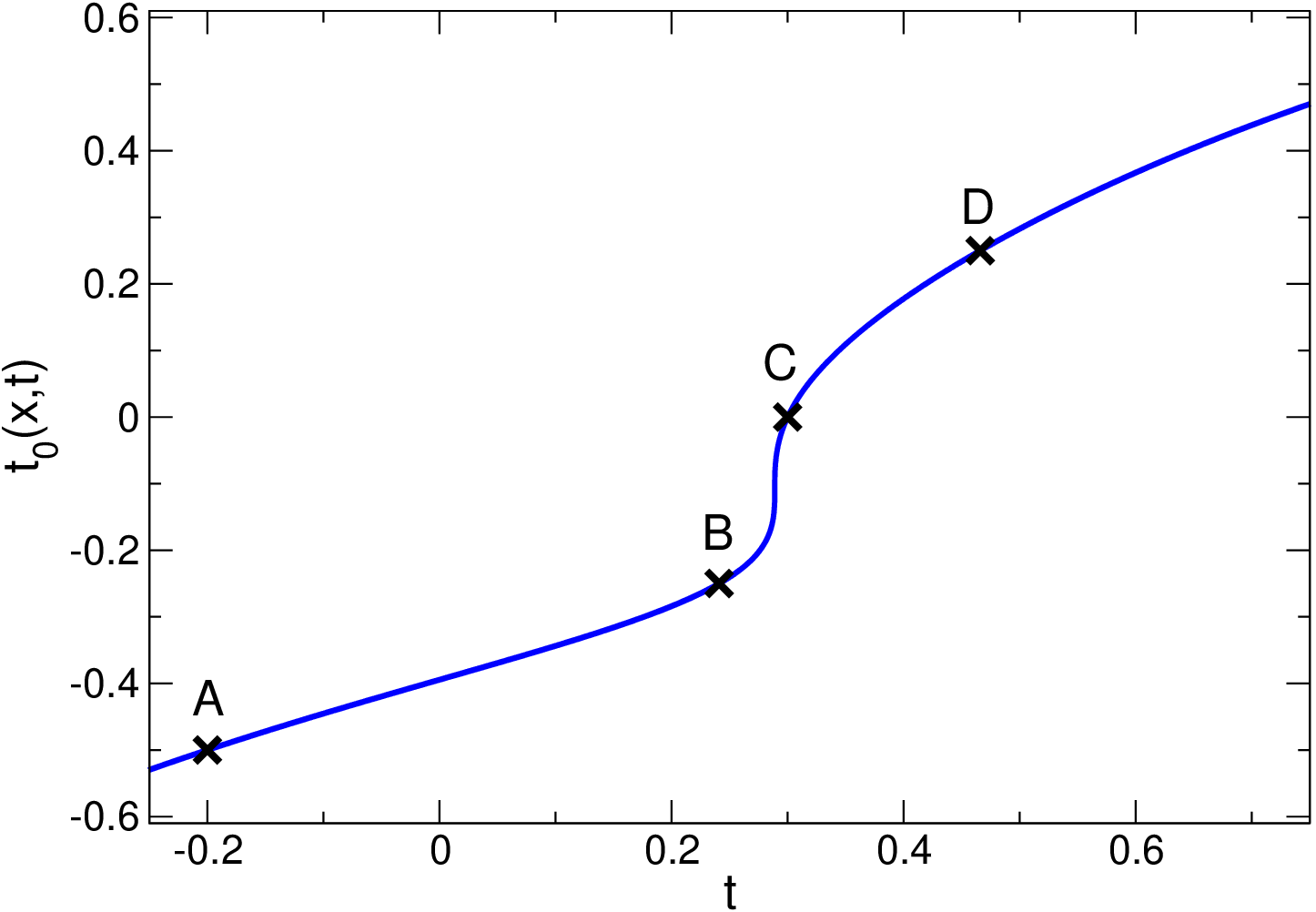}
  \caption{Mapping $t_0(x,t)$ from Eulerian to
    Lagrangian coordinates
    at $x=0.3$ for one period of $t$.}
  \label{fig:t0_map}
\end{figure}

The density
asymmetry about the peak also comes from the $1/u$ nonlinearity. Since
the compression is
enhanced in B--C relative to C--D, region B--C becomes smaller than
C--D, as seen in Fig.~\ref{fig:v_j_d_vt}. The results in
Fig.~\ref{fig:v_j_d_vt} are the solutions in Eulerian coordinates
which are obtained by composing the Lagrangian solutions in
Fig.~\ref{fig:v_j_d_vt0} with the map $t_0(x,t)$, i.e., the mapping
that determines which fluid element $t_0$ arrives at a given $(x,t)$,
shown in Fig.~\ref{fig:t0_map}.
Indeed while the maximum density fluid element lies nearly half way
between points B and C at the input (see
Fig.~\ref{fig:v_j_d_vt0}\footnote{Since the velocity of a fluid
  element does not change in the force-free drifting beam
  $u(x,t_0) = u(0,t_0)$. Hence $u(x,t_0)$ in Fig.~\ref{fig:v_j_d_vt0}
  can be used to determine the phase position of fluid element $t_0$
  with respect to the velocity modulation
  at $x=0$, even though the figure caption indicates that the
  quantities are evaluated at $x=0.3$.}) it is very near to point C by
$x=0.3$, the
point about which the linear density is maximum, as seen in
Fig.~\ref{fig:v_j_d_vt}. Because B--C is much more compressed than
C--D, loosely speaking, the regions that neighbor the region of
maximum density are a stretched region A--B earlier in time and the
compressed region C--D along with the stretched region D--A (periodic
waveform) later in time. Furthermore, the stretched region A--B has
enhanced stretching
over region D--A, contributing to the higher
density behind (in time) the peak.

In sum, the above describes how phase distortion of a force-free
drifting beam is a manifestation primarily of the inverse dependence
of the arrival time $t(x,t_0)$ on velocity $u$, but also depends on
the nonlinearity of the velocity modulation.
To prove that the ``dominant'' cause for phase distortion is
the $1/u$ nonlinearity we do
the following calculation. First, for a small velocity modulation
($\epsilon$ small) we can linearize the $1/u$ nonlinearity to get
\begin{eqnarray}
  \label{eq:approx_t}
  t(x,t_0) & \approx & t_0 + x (1-\epsilon \sin \omega t_0),\\
  \label{eq:approx_jac}
  \pdifftz{t} & \approx & 1 - x \epsilon \omega \cos \omega t_0.
\end{eqnarray}
For this arrival time function the location of maximum density
(minimum Jacobian) is the same as for the linear solution, point C in
Fig.~\ref{fig:v_j_d_vt}. However, with this approximation the problem
does not
immediately reduce to the linear solution since
the characteristics do not all have the same slope.\footnote{To get to
  the solution of the linear problem one can substitute
  Eq.~(\ref{eq:approx_jac}) into Eq.~(\ref{eq:lagrangian_continuity}),
  use that $\epsilon$ is small to move the denominator to the
  numerator with a sign change on the cosine term,
  and use $t_0 = t - x$ which can be obtained by taking $\epsilon = 0$
  in Eq.~(\ref{eq:approx_t}).} In this limit there is no phase distortion
since Eq.~(\ref{eq:approx_jac}) is an even function about the Jacobian
minimum, and hence the density is an even function about the density
peak. Furthermore, we could replace the modulation by any
periodic function that is odd in $t_0$ (plus or minus an arbitrary
time shift) for a period and get the same result. Thus, by {\em
  linearizing} the $1/u$ nonlinearity and {\em not linearizing} the
velocity modulation the phase distortion is removed. This confirms
that the $1/u$ nonlinearity is the dominant cause of phase
distortion.\footnote{One might be tempted to linearize the modulation
  without linearizing the $1/u$ nonlinearity for a complementary
  view. However, it is not
  possible to linearize the modulation and maintain its periodicity
  since a periodic function is nonlinear in its argument.}

\begin{figure}[htbp]
  \centering
  \includegraphics*[scale=\scaleval]{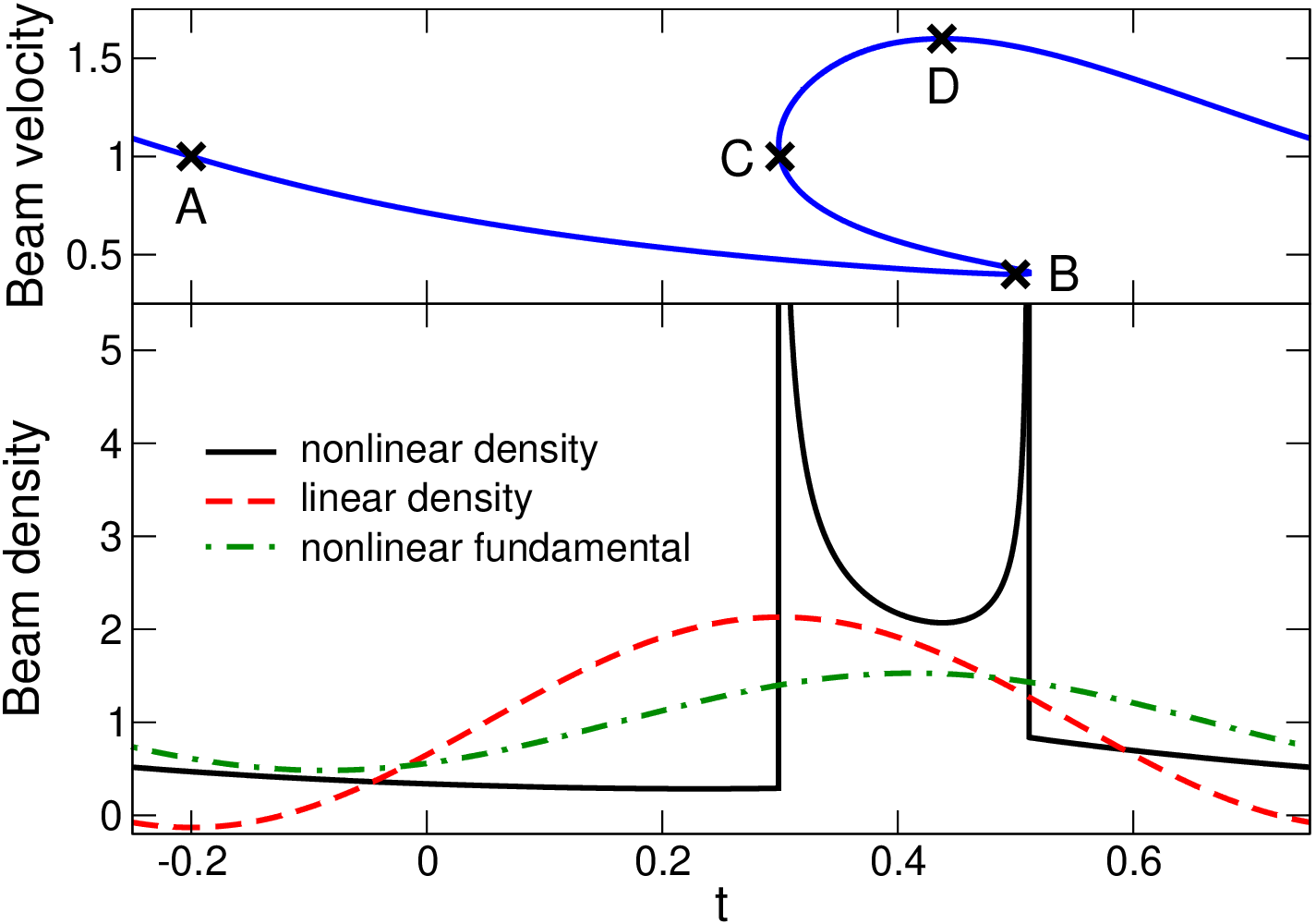}
  \caption{Nonlinear beam density, linear beam
    density, fundamental
    component of the nonlinear beam density, and nonlinear beam
    velocity for an input such that electron overtaking occurs.}
  \label{fig:overtake}
\end{figure}

Up to now we have restricted the input to a level such that electron
trajectories do not cross. The reason for such a
restriction was to ensure that the phase distortion physics was not
potentially clouded by electron overtaking physics. In fact, no such
restriction was necessary, and the principle of phase distortion is
the same even when the input is set such that electron trajectories
cross. In Fig.~\ref{fig:overtake} we show the beam velocity, nonlinear
beam density, linear beam density, and the fundamental component of
the nonlinear beam density for an input of $\epsilon = 0.6$. The
velocity solution and the two peaked structure of the density confirm
the multi-valued nature of the solutions. From the fluid element
labels we see that region B--D is in the multi-phase region of the
density, and that the higher density behind the peak is again because
region D--A (or region B--A outside the multi-phase zone) is not
stretched to the extent that region A--B (or region A--C outside the
multi-phase zone) is.

\begin{figure}[htbp]
  \centering
  \includegraphics*[scale=\scaleval]{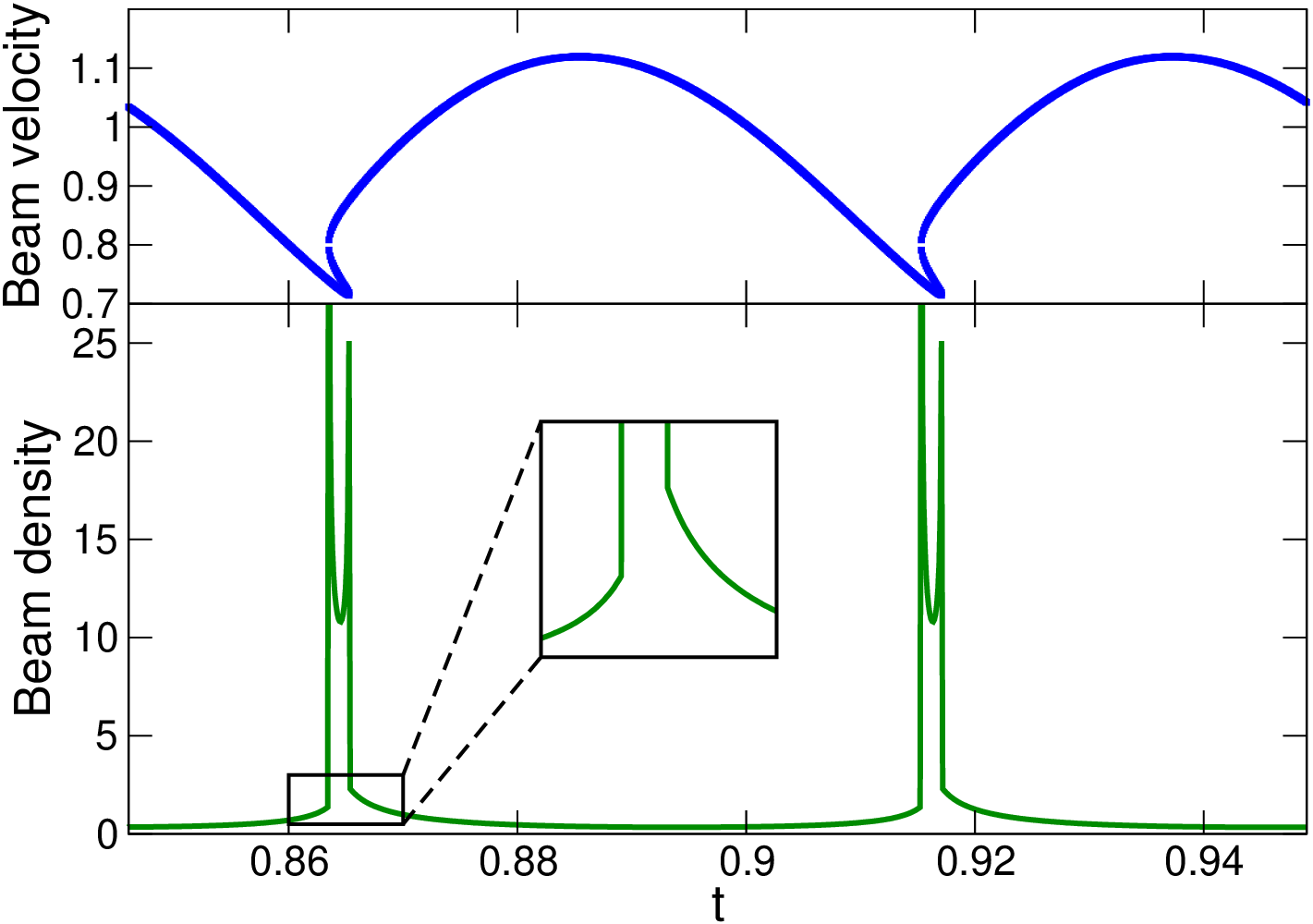}
  \caption{Beam density and beam velocity versus time
    for two periods
    from a Lagrangian TWT calculation
    accounting for space charge and circuit fields. The expanded view
    of the beam density
    is to highlight the density asymmetry about the peak.}
  \label{fig:twt}
\end{figure}

Even though the preceding analysis considered only a force-free
drifting beam, we claim that phase distortion mechanism identified
holds more generally for klystrons when space charge forces are
considered, and in traveling wave tubes (TWTs). The reason
is again the inverse dependence of arrival time on velocity. That is,
electrons slowed down from the dc beam velocity will have enhanced
stretching or compression over those that are sped up from the dc beam
velocity.
Although we do not provide an exhaustive analysis, for illustration 
we consider the electron beam density versus time from a
Lagrangian TWT model~\cite{wohlbier:ecwbtwt05} in
Fig.~\ref{fig:twt}.
Even though electrons experience forces due to space charge and
circuit fields in a TWT, as can be inferred
by comparing the beam velocities of
Figs.~\ref{fig:overtake} and~\ref{fig:twt}, the cause of
phase distortion is the same. Electrons slowed down from the dc beam
velocity will spread more in time than electrons sped up from the dc
beam velocity. The result is that the density wave form has a
relatively higher density behind the peak than in front of it, as seen
in Fig.~\ref{fig:twt}.

In Refs.~\cite{wohlbier:ampm04, wohlbier:nscw05} we described phase
distortion from a frequency domain perspective
as a ``self-intermodulation'' process, whereby harmonic
distortions mix with the fundamental to produce distortions at the
fundamental. Below we outline this view of phase distortion so that
it may be compared to the time domain view given
above.

If we express the velocity
$u$ and density $\rho$ with Fourier series (for periodic inputs),
then the fundamental frequency components of
nonlinear products of $\rho$ and $u$ are seen to come from mixing of
the second harmonic with the fundamental frequency, and mixing of higher
order harmonics. In particular,
if we define state variable envelopes as in
Refs.~\cite{wohlbier:ampm04, wohlbier:nscw05}, e.g.,
\begin{eqnarray}
  u(x,t) & = & \sum_{m=-\infty}^\infty \uhm(x) e^{i m \omega (x-t)},
\end{eqnarray}
then the continuity equation for the force-free drifting beam gives
\begin{eqnarray}
  \label{eq:mixing}
  \odiffx{\rhl} & = & - i \ell \omega \uhn + i \sum_{m,n \ne 0 \atop
    m+n = \ell} m \omega \uhm
  \uhn \nonumber\\
  && \mbox{} - i\ell\omega \sum_{m,n \ne 0 \atop m+n=\ell} \uhm \rhn
\end{eqnarray}
where the approximations used in
Refs.~\cite{wohlbier:ampm04, wohlbier:nscw05} have been made.
From Eq.~(\ref{eq:mixing}) products of
frequencies such as $(m,n)=(-1,2), (2,-1), (-2,3), (3,-2),$ etc.~are
seen to influence
the fundamental frequency $\ell = 1$. In
Refs.~\cite{wohlbier:ampm04, wohlbier:nscw05}
we considered $(m,n)=(-1,2), (2,-1)$
(third order intermodulation, ``$3\,$IM'') and $(m,n)=(-2,3), (3,-2)$
($5\,$IM) contributions to phase distortion in linear beam devices.

Equation~(\ref{eq:mixing}) together with the
corresponding nonlinear envelope equation for the velocity has an
analytic solution that may be expressed in an infinite series of
complex exponentials. The first terms in the series correspond to the
linear solution as seen in Fig.~\ref{fig:nl_l_fc}. The linear terms
are used in the equation for the second harmonic envelopes
to produce the first terms in the
series solution at the second harmonic. The second harmonic terms are
then combined with the linear terms from the fundamental solution to
produce the next set of terms at the fundamental. These complex
exponentials add to the linear density solution and produce a phase
shift (distortion) in the density solution. As it turns out, this
process of generating harmonics and then
additional terms at the fundamental must be continued to get an
accurate representation of the nonlinear phase shifted density
seen in Fig.~\ref{fig:nl_l_fc}. For the example in
Fig.~\ref{fig:nl_l_fc} we found that terms higher than $11\,$IM were
required to adequately approximate the nonlinear density, whereas in
Refs.~\cite{wohlbier:ampm04, wohlbier:nscw05} we found that $5\,$IMs
were sufficient to predict the phase distortion. The difference between
the two cases is that the case in
Fig.~\ref{fig:nl_l_fc} has a much larger relative input modulation.

We have identified the mechanism for phase distortion in the simplest
system that pertains to linear beam vacuum devices, the force-free
drifting electron beam. The dominant cause of phase distortion is
shown to be the inverse dependence of arrival time for an electron on
its velocity, i.e., the ``$1/u$ nonlinearity,'' and that a secondary
cause of phase distortion is the nonlinearity of the velocity
modulation. Although we show this to be the case for the force-free
drifting beam, we claim that the $1/u$ nonlinearity is also the cause
of phase distortion in other linear beam devices, such as the
traveling wave tube. That is, even though electrons are experiencing
forces from the circuit and space charge fields, the fact remains that
charges slowed down from the dc beam velocity will stretch and
compress more than charges sped up from the dc beam
velocity. Results from a traveling wave tube calculation are given
that show the same characteristic asymmetric density modulation that
is seen in the force-free drifting beam. We also show that the $1/u$
nonlinearity is the cause of phase distortion regardless
of whether drive levels are strong enough such that electron
overtaking occurs.

The identification of the mechanism for phase distortion suggests how
inputs might be
tailored to ameliorate phase distortion. The obvious candidates would
involve somehow reducing the amplitude of the velocity modulation on
the negative half cycle to lessen the effect of the ``enhanced
stretching,'' or to provide a density modulation $180^\circ$ out of
phase from the velocity modulation so that the ``enhanced stretching''
before the peak starts from a higher density value relative to the
density behind the peak,
and would be balanced about the density
peak at the output. The latter scheme may be facilitated by cold
cathode technology~\cite{whaley:edegtwt02}. It is anticipated that
either of these schemes, and potentially any scheme, may come with an
associated gain compression, as in Ref.~\cite{wohlbier:ampm04} where
harmonic injection was used to set the fundamental output phase at a
cost of reducing output power. It is also possible that further study
may show that any tailoring of the velocity modulation may just be a
manifestation of harmonic injection, proving the usefulness of the
complementary views of phase distortion given in this note and in
Refs.~\cite{wohlbier:ampm04,wohlbier:nscw05}.

\section*{Acknowledgments}
The author would like to thank Professor J.H.~Booske for a critical
review of the manuscript.

J.G.~W\"ohlbier was funded by the U.S.~Department of Energy
at Los Alamos National Laboratory, Threat Reduction Directorate, as an
Agnew National Security Postdoctoral Fellow.


\end{document}